\documentstyle[twocolumn,prb,aps,psfig]{revtex}

\title
{
Comment on "On the ${\mathbf T_S}$-Anomaly in Betaine Calcium Chloride Dihydrate"
}

\author
{ 
Boris Neubert$^{1}$ and Michel Pleimling$^{2}$
}
\address
{
$^1$ Departamento de F\'{\i}sica de la Materia Condensada,
  Facultad de Ciencias,
  Universidad del Pa\'{\i}s Vasco,
  Apartado 644, Bilbao, Spain\\
$^{2}$ Institut f\"ur Theoretische Physik, Technische Hochschule,
D--52056 Aachen, Germany
}

\begin{document}
\sloppy
\maketitle
Recently, Hlinka and Ishibashi\cite{ref:1}
discussed the $T_S$-anomaly\cite{ref:2}
in betaine calcium chloride dihydrate (BCCD) inside the fourfold modulated
phase in the framework of the phenomenological Landau theory. This anomaly
was interpreted as the signature of a phase transition from a high temperature
phase with space group P2$_1$ca to a low temperature phase 
with a {\it different}
space group. The same authors also proposed\cite{ref:3} a modified DIFFOUR
model for the description of a transition between two commensurate phases
with the same modulation period.

In the following, we want to comment on the shortcomings of the proposed
approach and discuss the $T_S$-anomaly in the framework of a microscopical
pseudo spin model based on a realistic description of BCCD in terms of
symmetry-adapted local modes (SALMs)\cite{ref:4,ref:5}.

BCCD is a unique material with more than twenty different commensurately (C)
and incommensurately (IC) modulated phases (for a recent review of the 
experimental situation, see reference\cite{ref:6}). The problems a Landau
description of that material encounters are well-known: for every
modulated phase a different free energy with different coefficients is
needed. A rather involved derivation of the theoretical temperature-pressure
phase diagram of BCCD based on a phenomenological approach is given in
reference\cite{ref:7}. The Landau free energy discussed by Hlinka and Ishibashi 
is tailored for the description of a successive ordering of two order parameters
and does not take into account the different modulated phases. 

The main problem of the approach of Hlinka and Ishibashi is their 
hypothesis that the symmetry
of the low-temperature fourfold phase differs from P2$_1$ca. In fact,
recent neutron scattering investigations\cite{ref:8} lead to the unambiguous
conclusion that the space group of the fourfold phase at 100 K, i.e.\
below $T_S \approx 115 K$, is P2$_1$ca, in accordance with earlier 
X-ray investigations\cite{ref:9}. Thus, the starting argument of
Hlinka and Ishibashi is in contradiction to the experimental
situation in BCCD.

A microscopic model suitable for a class of uniaxially structurally
modulated compounds can be derived as follows.
Due to the pseudo symmetry of half a lattice constant along ${\mathbf c}$, BCCD can be considered 
to be built up of successive layers of half cells perpendicular to ${\mathbf c}$.
Applying the method described in reference\cite{ref:4} to BCCD, the symmetry-breaking atomic
displacements occuring below the transition from the unmodulated high-temperature phase to the 
modulated phases are expanded in terms of a
localized symmetry-adapted basis set; the respective coordinates (mode amplitudes) 
are projected onto two-valued
pseudo spin variables leading to symmetry-based pseudo spin models. 
Every basis vector (SALM)
describes a collective displacement of all atoms in a half cell.
One obtains two relevant SALMs for the front and two for the back half cell. Adequate
superpositions of these describe C and IC structural modulations transforming 
according to the irreducible 
representations $\Lambda_2$ and $\Lambda_3$ of the group of the wave 
vector ${\mathbf q}= q {\mathbf c}^*$.
The pseudo spins $\tau$ and $\sigma$ represent the signs of the amplitudes of
the two relevant SALMs.
Taking the transformation properties of the SALMs into account\cite{ref:4},  
the Hamiltonian 
%\begin{eqnarray}
%{\mathcal H} = K \sum\limits_{ijk} \tau_{ijk} \tau_{ij(k+1)} + L \sum\limits_{ijk}
%\sigma_{ijk} \sigma_{ij(k+1)} \nonumber\\
%+ \frac{M}{2} \sum\limits_{ijk} \left(
%\sigma_{ijk} \tau_{ij(k+1)} - \tau_{ijk} \sigma_{ij(k+1)} \right)  \nonumber\\
%+ J \sum\limits_{ijk} \tau_{ijk} \left( \tau_{(i+1)jk} + \tau_{i(j+1)k}
%\right)  \nonumber\\
%+ J' \sum\limits_{ijk} \sigma_{ijk} \left( \sigma_{(i+1)jk} +
%\sigma_{i(j+1)k} \right)
%\label{equation:dis}
%\end{eqnarray}
of the Double Ising Spin (DIS) model is obtained\cite{ref:10}. 
This model is characterized by the competition between symmetric $\tau$-$\tau$
and $\sigma$-$\sigma$ interactions on the one hand and an antisymmetric
coupling between $\tau$ and $\sigma$ on the other hand, which leads to
frustration effects and hence to modulated structures. 

One interesting feature of the DIS model is the existence of phase transitions
between phases with equal wave numbers but different pseudo spin configurations
\cite{ref:11}. The character of these `internal' transitions has been
investigated by different methods: in mean field approximation,
with the mean field transfer matrix method, and in Monte Carlo 
simulations\cite{ref:12}.
The structural changes at the `internal' transitions are characterized by 
different strengths of harmonics in a Fourier analysis of the spatial
modulation. 
In the case of the `internal' transition from the high temperature to
the low temperature modification of the fourfold phase, 
the DIS model predicts a discontinuous change of the first and 
third order harmonics (all other harmonics being zero) towards a more anharmonic 
structure with both modifications exhibiting the {\it same} space group symmetries.
Neutron scattering studies\cite{ref:8} revealed a strong third harmonic in the 
displacements described by the $\sigma$-profile, which is in good 
agreement with the structure of the
low temperature modification obtained from the DIS model.

An interesting by-product of
our model calculations is the tentative explanation of the seemingly contradictory
observation of a more sinusoidal
modulation by X-ray techniques\cite{ref:9} at $T=90K$,
which can be well described as the high temperature modification
in the DIS model. Irradiation damage results in
(positive) plastic strains, that shift the boundaries
in the $p$-$T$-phase diagram to lower temperatures. 
In spite of the greater slope of the $T_S$-line, this shift is stronger for $T_S$
than for the other phase boundaries \cite{ref:2}.
Thus, X-ray induced defects 
might have opened up a window for the observation of the 
high temperature modification of the fourfold phase reported
in reference\cite{ref:9}.
The argumentation of Hlinka and Ishibashi concerning the defect influence on $T_S$
is incomplete since the shift of the other phase boundaries is disregarded.
In a more detailed discussion along the lines
given in section VIII of reference\cite{ref:5}, the different shifts
of $T_S$ and the other transition temperatures
can be consistently explained by observing that the 
internal phase transitions in the DIS model exhibit a different sensitivity on small
changes of the couplings (due to the defect-induced lattice distortion)
than the other phase boundaries.

The symmetry-based DIS model is suitable for the
explanation even of detailed characteristics of complex modulated real systems.
The $T_S$-anomaly in BCCD and its characteristic properties can be 
interpreted as a realization of the phase transitions between structures with the same 
wave number and symmetries but different atomic configurations:
the respective phases differ significantly
in the amplitudes of the harmonics in a Fourier expansion of the displacements.
The two experimentally determined structures agree fairly well 
with our calculated data.

One of the authors (B.N.) wishes to thank DAAD (Deutscher Akademischer
Austauschdienst) for financial support.


\begin{thebibliography}{99}
\bibitem{ref:1} J.~Hlinka and Y.~Ishibashi: J. Phys. Soc. Jpn. {\bf 67}
(1998) 495.
\bibitem{ref:2} M.~Le~Maire, R.~Straub and G.~Schaack: Phys. Rev. {\bf B 56}
(1997) 134.
\bibitem{ref:3} Y.~Ishibashi and J.~Hlinka: J. Phys. Soc. Jpn. {\bf 67}
(1998) 27.
\bibitem{ref:4} B.~Neubert, M.~Pleimling and R.~Siems: J. Kor. Phys. Soc.
{\bf 32} (1998) S36.
\bibitem{ref:5} B.~Neubert, M.~Pleimling and R.~Siems: Ferroelectrics 
{\bf 208-209} (1998) 141.
\bibitem{ref:6} G.~Schaack and M.~Le~Maire: Ferroelectrics (1998)
in print.
\bibitem{ref:7} 
D.~G.~Sannikov and G.~Schaack: J. Phys.: Condens. Matter {\bf 10} (1998)
1803.
\bibitem{ref:8} O.~Hernandez, M.~Quilichini, A.~Cousson, W.~Paulus, J.-M.~Kiat,
A.~Goukassov, J.~M.~Ezpeleta, F.~J.~Z\'{u}\~{n}iga, J.~M.~P\'{e}rez-Mato,
M.~Du\v{s}ek and V.~Pet\v{r}\'{i}\v{c}ek: in {\it Proceedings of Aperiodic 1997}
(1998) in press; O.~Hernandez: PhD thesis (Paris, France) (1997).
\bibitem{ref:9} J.~M.~Ezpeleta, J.~M.~P\'{e}rez-Mato, W.~Paciorek
and T.~Breczewski: Acta Cryst. B {\bf 48} (1992) 261.
\bibitem{ref:10} M.~Pleimling, B.~Neubert and R.~Siems: Z. Phys. B {\bf 104}
(1997) 125.
\bibitem{ref:11} B.~Neubert, M.~Pleimling and R.~Siems: 
international seminary 'Inkommensurable Strukturen', 
Erbenhausen/Germany (1995).
\bibitem{ref:12} B.~Neubert, M.~Pleimling and R.~Siems: cond-mat/9803252.
\end{thebibliography}
\end{document}